\RequirePackage{fix-cm}
\documentclass[twocolumn,epjc3]{svjour3}  
\smartqed
\usepackage[T1]{fontenc}
\usepackage{amsmath,amssymb}
\usepackage{graphicx}
\usepackage{hyperref}
\usepackage{cite}
\usepackage{xcolor}
\usepackage{amsmath}
\usepackage{placeins}
\usepackage{graphicx}
\usepackage{subcaption}

\begin{document}
\title{Pair Production of Singlet Vector-Like B Quarks in Boosted bZ Final States at a Muon Collider}
\author{Eda Alici}
\institute{
Department of Physics, Zonguldak Bulent Ecevit University,
Zonguldak, Turkey
}
\date{Received: date / Accepted: date}
\maketitle
\begin{abstract}
We investigate the pair production of singlet vector-like bottom quarks at a future
$10~\mathrm{TeV}$ muon collider in the boosted $Zb$ final-state topology. The signal
process is considered as
$\mu^{+}\mu^{-}\to B\bar{B}\to (Zb)(Z\bar{b})$,
with one $Z$ boson decaying leptonically and the other hadronically, leading to a
final state with an opposite-sign same-flavour dilepton pair, multiple jets, and two
$b$-tagged jets. Signal and Standard Model background events are generated at
leading order with MadGraph5\_\allowbreak aMC@NLO and subsequently interfaced with
\textsc{Pythia}~8 for parton showering and hadronization. Detector effects are
simulated with \textsc{Delphes} using a muon-collider detector configuration. A
cut-based analysis exploiting boosted kinematics, including hard leading-$b$-jet
transverse momentum, small dilepton angular separation, and large hadronic activity,
is performed to suppress the dominant $ZZjj$, $ZZb\bar{b}$, and $t\bar{t}$
backgrounds. Statistical sensitivities are evaluated with the Asimov approximation,
including both zero and $10\%$ background systematic uncertainty scenarios. For the
singlet benchmark $\mathrm{BR}(B\to bZ)=0.25$ and an integrated luminosity of
$10~\mathrm{ab}^{-1}$, a $5\sigma$ discovery reach up to
$M_B\simeq 4.3~\mathrm{TeV}$ is obtained, reduced slightly to
$M_B\simeq 4.2~\mathrm{TeV}$ when systematic effects are included. These results
demonstrate that a $10~\mathrm{TeV}$ muon collider provides a powerful and clean
environment for probing vector-like $B$ quarks well beyond present LHC limits.
\end{abstract}
\keywords{VLQ-B \and muon collider \and BSM physics}
\section{Introduction}
\label{sec:intro}
The Standard Model (SM) of particle physics provides a well-established theoretical framework that has been tested with high precision in a wide range of collider experiments~\cite{Glashow1961,Weinberg1967,Salam1968,Gaillard1999}. Despite its remarkable success, the SM is widely believed to be an incomplete description of nature, as it fails to address several fundamental questions such as the hierarchy problem~\cite{Hierarchy_Susskind1979,Hierarchy_tHooft1980}, the origin of dark matter~\cite{Bertone2005}, and the observed matter--antimatter asymmetry in the universe~\cite{Sakharov1967}. These theoretical shortcomings have motivated extensive searches for physics beyond the SM (BSM) at current and future high-energy collider experiments.

One of the compelling extensions of the SM involves the introduction of vector-like quarks (VLQs), exotic spin-$1/2$ fermions whose left- and right-handed chiral components transform identically under the SM gauge group $\mathrm{SU}(3)_C \times \mathrm{SU}(2)_L \times \mathrm{U}(1)_Y$~\cite{VLQ_review1,VLQB_pp_Zb1,Alves:2024vlq,Okada:2012gy}. In contrast to chiral SM quarks, VLQs can possess gauge-invariant vector-like mass terms that are independent of electroweak symmetry breaking. As a result, their masses can be naturally large without requiring sizeable Yukawa couplings to the Higgs field. This feature makes VLQs theoretically well motivated in scenarios addressing the hierarchy problem, where they may contribute to the cancellation of quadratic divergences in the Higgs boson mass through radiative corrections~\cite{DeSimone2013TopPartner,Angelescu2016VLQHiggs}. VLQs arise in a wide range of BSM scenarios, including composite Higgs models, extra-dimensional theories, and little Higgs models~\cite{Chala2018VLQFuture,Atre2011SingleVLQ}. This broad theoretical motivation has led to extensive searches for VLQs at current and future colliders.

The  phenomenology of VLQs is not universal; it depends on the electroweak representation, the mixing pattern with SM quarks, and the dominant decay channels. VLQs may appear as singlets, doublets, or triplets under $\mathrm{SU}(2)_L$, leading to different charge assignments and experimental signatures. Besides the top-like partner $T$ with electric charge $+2/3$ and the bottom-like partner $B$ with electric charge $-1/3$, exotic states such as $X$ and $Y$, carrying electric charges $+5/3$ and $-4/3$, respectively, can also occur in extended VLQ multiplets~\cite{VLQB_pp_Zb1,Banerjee2024VLQStatus}. A broad range of studies has explored these possibilities for top-like, bottom-like, and exotic-charge VLQs at the present and future colliders~\cite{Aguilar-Saavedra2013,Chala2018VLQFuture,Yang2019SingleVLQ,Angelescu2016VLQHiggs,Liu2016SingleB,Chen2016SingleB,Carvalho2018VLQ,Banerjee2024TripletVLQ,Bardhan2023MLVLQ,Barducci2017VLQdiscrimination,Canbay2023SingleVLQ,Han2023EPJC846,Banerjee2024VLQStatus,Liu2024VLQX,Zhang2024VLQX_FCCeh,Yang2021VLQTProd,Cetinkaya2021SingleVLQ,Zhou2020NewDecayVLQ,Liu2017SingleTgamma,Lv2022MuonVLQ,Qin2022VLQT_CLIC,Han2025MuonProtonVLQ,Gong2019SingleB,Yang:2024VLQBhadronic,Han2022SingleVLQB,Han2022NuPhB975,ZengPanZhang2023,Gong2020SingleVLQB,Shang2022VLQB,Han2022VLQCLIC,Han2024PairVLQB,Yang2025CLICPairB,Benbrik2025VLQBounds}.

Among the possible VLQ representations, vector-like bottom-type quarks (hereafter denoted $B$), carrying electric charge $-1/3$, provide a particularly relevant benchmark for collider searches. These states can mix with the SM bottom quark, thereby acquiring couplings to the $W$, $Z$, and Higgs bosons and giving rise to the characteristic decay modes $B\to Wt$, $B\to bZ$, and $B\to bH$~\cite{VLQB_pp_Zb1,Angelescu2016VLQHiggs,Bardhan2023MLVLQ,Han2023EPJC846,Gong2019SingleB,Yang:2024VLQBhadronic,Han2022SingleVLQB,Han2022NuPhB975,ZengPanZhang2023,Gong2020SingleVLQB,Shang2022VLQB,Han2022VLQCLIC,Han2024PairVLQB,Yang2025CLICPairB}. These channels produce distinctive multi-object final states that can be exploited for discovery at hadron and lepton colliders alike. In the limit of large $B$ mass, the Goldstone boson equivalence theorem predicts that the partial widths into the longitudinal components of the $W$ and $Z$ bosons and into the physical Higgs boson become equal, yielding the asymptotic branching ratio $\mathrm{Br}(B \to Wt) : \mathrm{Br}(B \to bZ) : \mathrm{Br}(B \to bH) \to 2:1:1$ for the singlet representation. This ratio serves as a standard benchmark in collider analyses and is adopted throughout the present work.

It is useful to place the present study in the context of existing experimental searches and phenomenological projections for vector-like $B$ quarks. On the experimental side, ATLAS and CMS analyses of LHC Run-2 data have placed pair-production exclusion limits on vector-like quarks in the TeV mass range~\cite{ATLAS2024VLQpair,CMS:2022VLQpair}, while single-production searches yield coupling-dependent limits~\cite{ATLAS:2023VLQB_single,CMS2024VLQBsingle}. A recent unified reinterpretation of these results, performed with the VLQBounds tool, confirms a singlet pair-production exclusion of $M_B \gtrsim 1.33$--$1.49~\mathrm{TeV}$ and extends the doublet limit to $M_B \gtrsim 1.52~\mathrm{TeV}$~\cite{Benbrik2025VLQBounds}.

Beyond the LHC, single production of the singlet VLQ-$B$ quark via the $B\to tW$ channel has been studied at the HL-LHC, HE-LHC, and a 100~TeV FCC-hh collider for a benchmark coupling $g^\ast=0.2$. Using detailed detector simulations and background analyses, Ref.~\cite{Han2023EPJC846} reported that the HL-LHC can exclude (discover) singlet VLQ-$B$ masses up to about $2.1~(1.9)~\mathrm{TeV}$, while the exclusion reach extends beyond $3~\mathrm{TeV}$ at the HE-LHC and beyond $5~\mathrm{TeV}$ at the FCC-hh. This illustrates the strong scaling of the single-production sensitivity with collider energy.

For lepton--hadron colliders, single production of the singlet VLQ-$B$ quark has been investigated at the LHeC with $\sqrt{s}=1.98~\mathrm{TeV}$. In the $B\to Wt$ channel, the semileptonic final state was found to offer the best sensitivity among the fully hadronic, fully leptonic, and semileptonic topologies~\cite{Gong2020SingleVLQB}. In the $B\to bZ$ channel, the achievable sensitivity at the LHeC and its higher-energy counterpart, the FCC-eh with $\sqrt{s}=5.29~\mathrm{TeV}$, was shown to improve significantly with increasing collision energy. For the benchmark scenarios considered in Ref.~\cite{Shang2022VLQB}, the accessible mass range extends up to about $1.9~\mathrm{TeV}$ at the LHeC and up to about $2.4$--$2.8~\mathrm{TeV}$ at the FCC-eh, depending on whether discovery or exclusion sensitivity is considered.

Furthermore, at electron--positron colliders, single production of the singlet VLQ-$B$ quark via $e^+e^- \to B\bar{b}$ with $B\to bZ$ has been examined at the 3~TeV CLIC. For an integrated luminosity of $\mathcal{L}=5~\mathrm{ab}^{-1}$, the reported singlet sensitivity reaches the mass range $M_B\simeq1.2$--$1.9~\mathrm{TeV}$ at the $5\sigma$ discovery level and extends up to about $M_B\simeq2.8~\mathrm{TeV}$ at the 95\% C.L. exclusion level, depending on the coupling strength~\cite{Han2022VLQCLIC}. Complementary studies have explored pair production of the singlet VLQ-$B$ quark through $e^+e^- \to B\bar{B}$ at the same CLIC energy, both in the $bZ$ channel~\cite{Han2024PairVLQB} and in fully hadronic final states, with a sensitivity extending up to approximately $M_B\lesssim1.5~\mathrm{TeV}$ for $\mathcal{L}=5~\mathrm{ab}^{-1}$~\cite{Yang2025CLICPairB}.

The existing limits and projections indicate that substantially higher centre-of-mass energies are required to probe vector-like $B$ quarks in the multi-TeV mass range, especially in pair-production channels. Future high-energy lepton colliders are therefore particularly attractive, as they combine a clean experimental environment with direct access to heavy new states. In this context, the muon collider has emerged as one of the most promising candidates for the next energy-frontier machine~\cite{Long2021Muon,MuonColliderForum2022}. Unlike electron--positron colliders, which suffer from severe synchrotron radiation losses at high energies, muon colliders can in principle reach multi-TeV centre-of-mass energies with relatively compact circular ring designs, owing to the muon's approximately 200 times greater mass compared to the electron. A muon collider operating at $\sqrt{s} = 10~\mathrm{TeV}$ would provide access to new particle masses well beyond the reach of the High-Luminosity LHC, while simultaneously retaining the theoretical cleanliness of leptonic initial states and the precision associated with a well-defined beam energy. Recent studies have shown that a multi-TeV muon collider would support a broad physics program, ranging from BSM searches to precision tests of the SM.

In this work, we study singlet vector-like $B$ quark pair production at a $\sqrt{s}=10~\mathrm{TeV}$ muon collider through the Drell--Yan-like process $\mu^+\mu^- \to B\bar{B}$, mediated by $s$-channel photon and $Z$-boson exchange. The high collision energy considered here allows the pair-production reach to extend to multi-TeV $B$ masses, while the clean lepton--antilepton initial state facilitates the reconstruction of the boosted $Zb$ final state. To quantify this potential,  we compute the pair-production cross section as a function of $M_B$ and focus on the decay chain $B\bar{B}\to (Zb)(Z\bar{b})$, with one $Z$ boson reconstructed leptonically and the other hadronically. Signal and background processes are evaluated at leading order (LO) using Monte Carlo event generators, and the discovery prospects are assessed through a cut-based analysis using the Asimov significance formula~\cite{Cowan2011Asymptotic}. Our results indicate that a $10~\mathrm{TeV}$ muon collider can probe vector-like $B$ quarks over a mass range well above current experimental bounds, and we provide quantitative estimates of the expected discovery reach as a function of the integrated luminosity.

The remainder of this paper is structured as follows. Section~\ref{sec:framework} establishes the theoretical framework, detailing the effective Lagrangian and the targeted signal production topologies. Section~\ref{sec:simulation} describes the Monte Carlo simulation chain, including the event generation and fast detector simulation setup. The relevant SM background processes, kinematic distributions, and the event selection strategy are presented in Section~\ref{sec:selection}. Section~\ref{sec:analysis} is dedicated to the statistical analysis and evaluates the projected discovery and evidence sensitivities at a 10~TeV muon collider. Finally, our main conclusions and final remarks are summarized in Section~\ref{sec:conclusion}.

\section{Theoretical Framework and Signal Topology}
\label{sec:framework}
In the present work, the singlet vector-like $B$ quark interactions are described within the framework of a simplified effective Lagrangian model~\cite{VLQ_review2,Buchkremer2013VLQUFO,VLQB_pp_Zb1}. The relevant interaction terms between the singlet VLQ-$B$ quark and the SM particles can be written in the following effective Lagrangian form:
\begin{equation}
\begin{aligned}
\mathcal{L}_{\mathrm{eff}} =
\frac{g g^\ast}{2}
\Bigg[&
\frac{1}{\sqrt{2}}\,
\bar{B}_L W_\mu^{+}\gamma^\mu t_L
+\frac{1}{2c_W}\,
\bar{B}_L Z_\mu\gamma^\mu b_L
\\
&-\frac{m_B}{2m_W}\,
\bar{B}_R H b_L
-\frac{m_b}{2m_W}\,
\bar{B}_L H b_R
\Bigg]
\\
&-\frac{e}{6c_W}\,
\bar{B}B_\mu\gamma^\mu B\, 
-\frac{e}{4s_W}\,
\bar{B}W_\mu^{3}\gamma^\mu B
+\mathrm{h.c.}
\end{aligned}
\label{eq:Leff}
\end{equation}
where $g$ denotes the weak $\mathrm{SU}(2)_L$ gauge coupling constant, $g^\ast$ is the effective VLQ--SM coupling parameter, and $s_W=\sin\theta_W$ and $c_W=\cos\theta_W$, with $\theta_W$ being the Weinberg angle. The quantities $m_B$ and $m_W$ denote the masses of the vector-like $B$ quark and the $W$ boson, respectively. The fields $W_\mu^-$, $Z_\mu$, and $H$ correspond to the $W^-$ boson, $Z$ boson, and physical Higgs boson fields, while $B_{L,R}$, $b_{L,R}$, and $t_L$ denote the relevant chiral quark fields. The term h.c. denotes the Hermitian conjugate.

In the present analysis, the signal process is based on the pair production of singlet VLQ-$B$ quarks at a high-energy muon collider. In this pair-production mechanism, the dependence of the signal cross-sections on the effective coupling parameter differs significantly from that of the single production channels. Unlike single production processes, the pair production mechanism of the singlet VLQ-$B$ quark at lepton colliders is largely insensitive to the effective coupling parameter $g^\ast$, since the production cross section is mainly determined by the VLQ mass and electroweak gauge interactions. In addition, the branching ratio pattern of the dominant decay channels ($\mathrm{BR}(B\to Wt) : \mathrm{BR}(B\to bZ) : \mathrm{BR}(B\to Hb) \simeq 2:1:1$), as predicted by the Goldstone boson equivalence theorem~\cite{He1992Equivalence,He1994Equivalence}, remains approximately unchanged in the simplified singlet pair-production scenario. However, the total decay width of the VLQ-$B$ quark is directly controlled by the coupling parameter $g^\ast$. Therefore, the benchmark value $g^\ast=0.1$ is adopted in this analysis in order to ensure that the decay width remains sufficiently small compared to the VLQ mass over the considered parameter space. In the considered parameter space, the decay width of the vector-like $B$ quark satisfies the condition $ \Gamma_B/M_B < 1\%$
ensuring the validity of the narrow-width approximation (NWA). Consequently, the production and decay stages of the signal process can be factorized consistently throughout the analysis~\cite{NWA_general1,NWA_general2,Kauer2013NWA}.

\begin{figure}[t]
  \centering
  \includegraphics[width=0.55\textwidth]{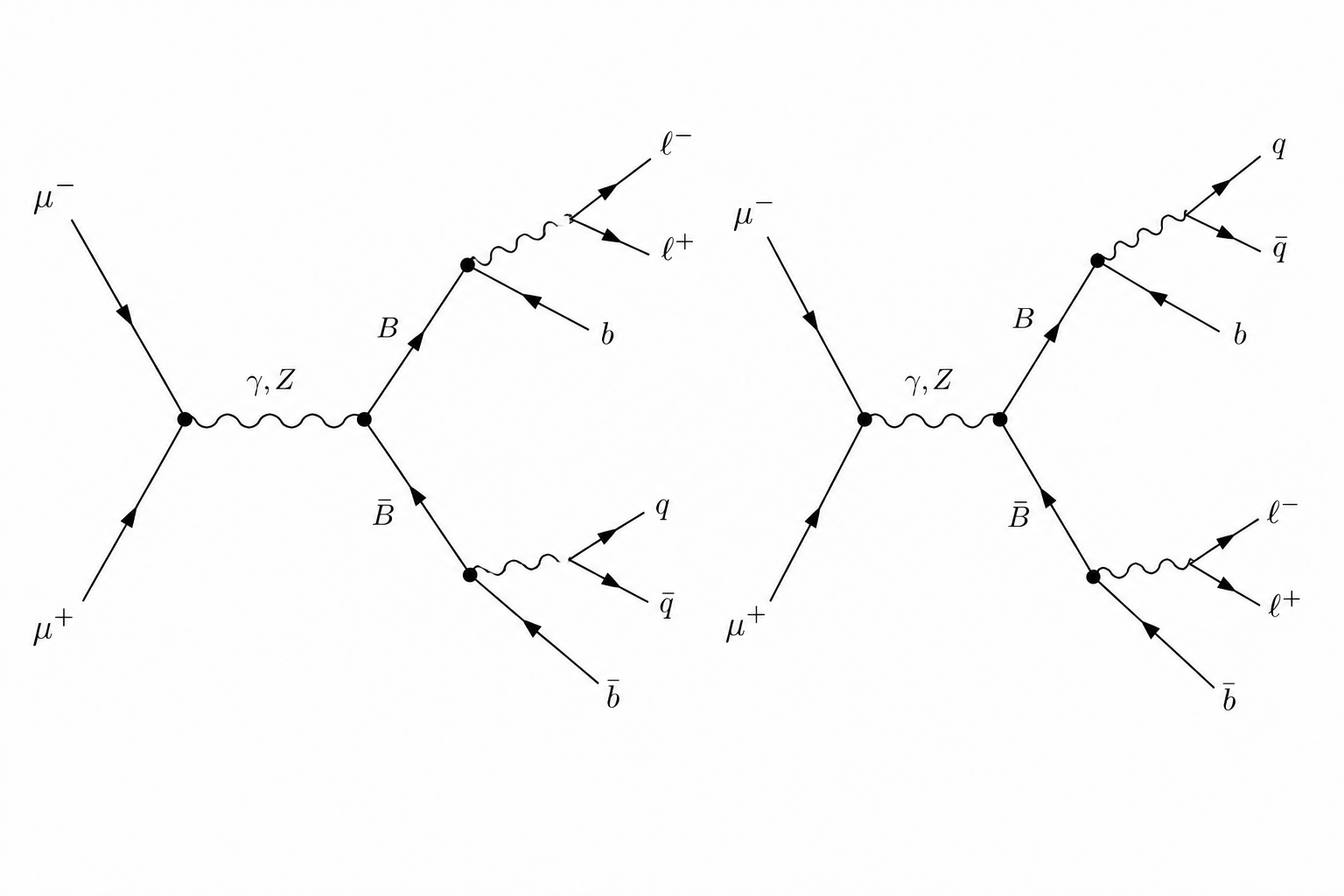}
  \caption{Representative tree-level Feynman diagrams for
  $\mu^+\mu^- \to B\bar{B}\to (Zb)(Z\bar{b})\to
  \ell^+\ell^- b\, jj\bar{b}$. The pair-produced VLQ-$B$ quarks are generated
  via $s$-channel $\gamma/Z$ exchange, and both permutations of the leptonic and
  hadronic $Z$-boson decays are included.}
  \label{fig:feynman_subprocess}
\end{figure}

The corresponding leading-order production subprocess is $\mu^+\mu^- \to B\bar{B}$. Both heavy quarks are assumed to decay via $B\to Zb$, and a mixed leptonic--hadronic topology is considered for the two resulting $Z$ bosons: one decays as $Z\to\ell^+\ell^-$ and the other as $Z\to jj$. The full signal process is therefore
\begin{equation}
\mu^+\mu^- \to B\bar{B} \to (Zb)(Z\bar{b}) \to \ell^+\ell^- b + jj\bar{b},
\end{equation}
yielding a final state with two opposite-sign same-flavour leptons, multiple jets, and two $b$-tagged jets. Representative tree-level Feynman diagrams for this signal topology are shown in Fig.~\ref{fig:feynman_subprocess}.

This final state provides a relatively clean experimental signature due to the leptonic reconstruction of the $Z$ boson, while the hadronic decay channel enhances the overall event reconstruction efficiency. Furthermore, the large mass of the vector-like quarks leads to energetic final-state objects and enhanced hadronic activity, providing a strong kinematic separation between the signal and the SM background processes. In the present analysis, the vector-like $B$ quark is treated within the NWA, since the total decay width remains much smaller than the resonance mass throughout the considered parameter space. Under this assumption, the production and decay stages can be factorized, allowing the total signal rate to be expressed in terms of the production cross section multiplied by the corresponding branching ratios~\cite{NWA_general1,NWA_general2,Kauer2013NWA}. Accordingly, under the NWA assumption, the total signal cross section for the process under consideration can be expressed as
\begin{multline}
\sigma(\mu^+\mu^- \rightarrow B\bar{B}
\rightarrow Zb\,Z\bar{b}
\rightarrow \ell^+\ell^- b\, jj\bar{b}) \\
= \sigma(\mu^+\mu^- \rightarrow B\bar{B})
\times \mathrm{BR}(B \rightarrow bZ)^2 \\
\times 2\,\times \mathrm{BR}(Z \rightarrow \ell^+\ell^-)
\times \mathrm{BR}(Z \rightarrow jj).
\label{eq:nwa_pairproduction}
\end{multline}
The combinatorial factor of two accounts for the two possible assignments in which either of the two $Z$ bosons from the $B\bar{B}$ decay chain decays leptonically, while the other decays hadronically.

\section{Monte Carlo Simulation and Detector Effects}
\label{sec:simulation}

The signal and SM background events are generated at leading order (LO) using \texttt{MadGraph5\_aMC@NLO v3.6.3}~\cite{Alwall2014MG5}. The interactions of the vector-like $B$ quark are implemented through the corresponding UFO model within the simplified model framework~\cite{Degrande2012UFO,Alloul2014FeynRules,VLQB_pp_Zb1,Buchkremer2013VLQUFO,VLQBsingletUFO}. The generated parton-level events are subsequently passed to \texttt{PYTHIA 8.245}~\cite{Sjostrand2015Pythia8} for parton showering and hadronization. Detector effects are simulated using the \texttt{Delphes 3.5.1} fast detector simulation package~\cite{deFavereau2014Delphes} employing the muon collider detector card. Jets are reconstructed using the Valencia Linear Collider (VLC) algorithm with a radius parameter of $R=0.5$~\cite{Boronat2014VLC}. A $b$-tagging efficiency of $70\%$ is assumed throughout the analysis. In order to preserve the efficiency of boosted dilepton final states, a loose lepton isolation criterion with $\Delta R = 0.05$ is adopted throughout the analysis. Finally, the event reconstruction, cut-flow analysis, and kinematic studies are performed within the \texttt{ROOT} analysis framework~\cite{ROOTFramework}.

\section{Event Selection and Kinematic Analysis}
\label{sec:selection}

\subsection{Standard Model Background Processes}

In order to enhance the experimental sensitivity of the signal process, the kinematical and topological properties of both the signal and the corresponding SM background processes are carefully investigated. Based on these characteristics, an event selection strategy is designed to maximize the signal significance while suppressing the background contributions.

For the final state topology containing two opposite-sign same-flavor leptons, multiple jets, and two b-tagged jets, the dominant SM background processes considered in this analysis are given by
\begin{equation}
\mu^+\mu^- \to ZZjj, \quad \mu^+\mu^- \to ZZb\bar{b}, \quad \mu^+\mu^- \to t\bar{t}.
\end{equation}

\subsection{Signal Optimization and Cut-Flow Analysis}

\begin{figure*}[t]
\centering
\begin{minipage}{0.48\textwidth}
    \centering
    \includegraphics[width=\textwidth]{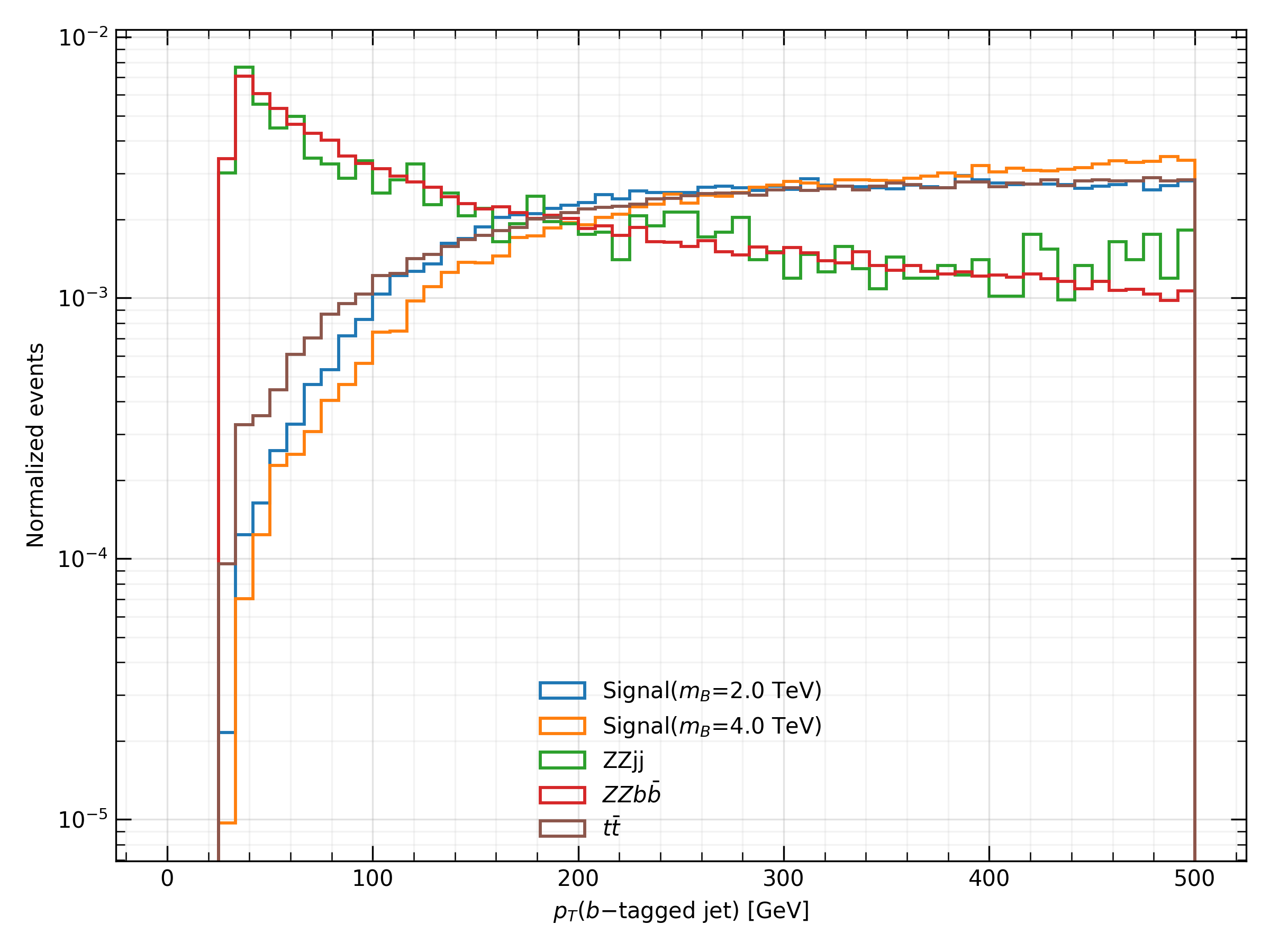}
    \par\small (a) Transverse momentum distribution of the leading $b$-tagged jet, $p_T(b_1)$
\end{minipage}
\hfill
\begin{minipage}{0.48\textwidth}
    \centering
    \includegraphics[width=\textwidth]{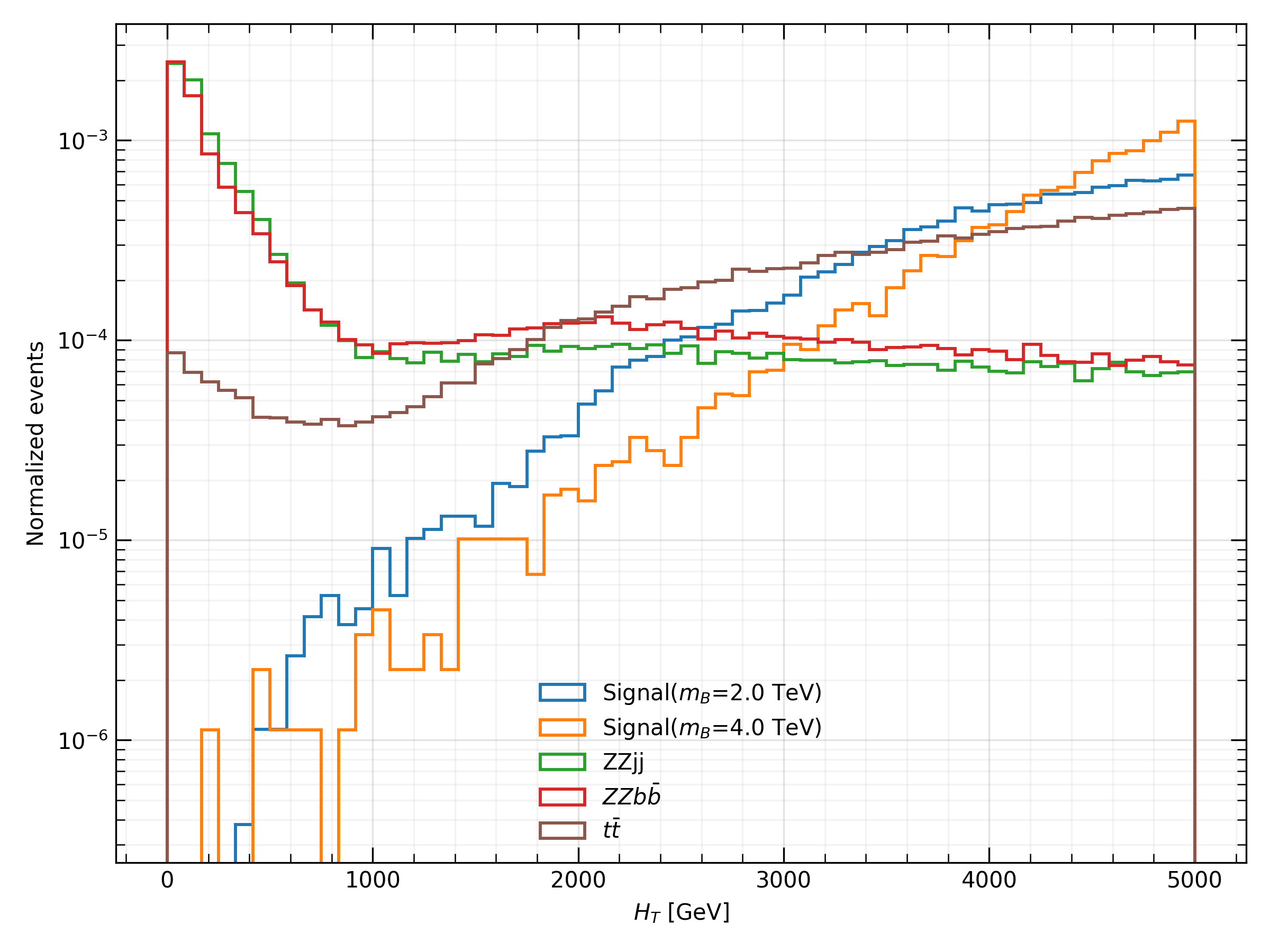}
    \par\small (b) Scalar sum of jet transverse momenta, $H_T$
\end{minipage}
\vspace{0.4cm}
\begin{minipage}{0.48\textwidth}
    \centering
    \includegraphics[width=\textwidth]{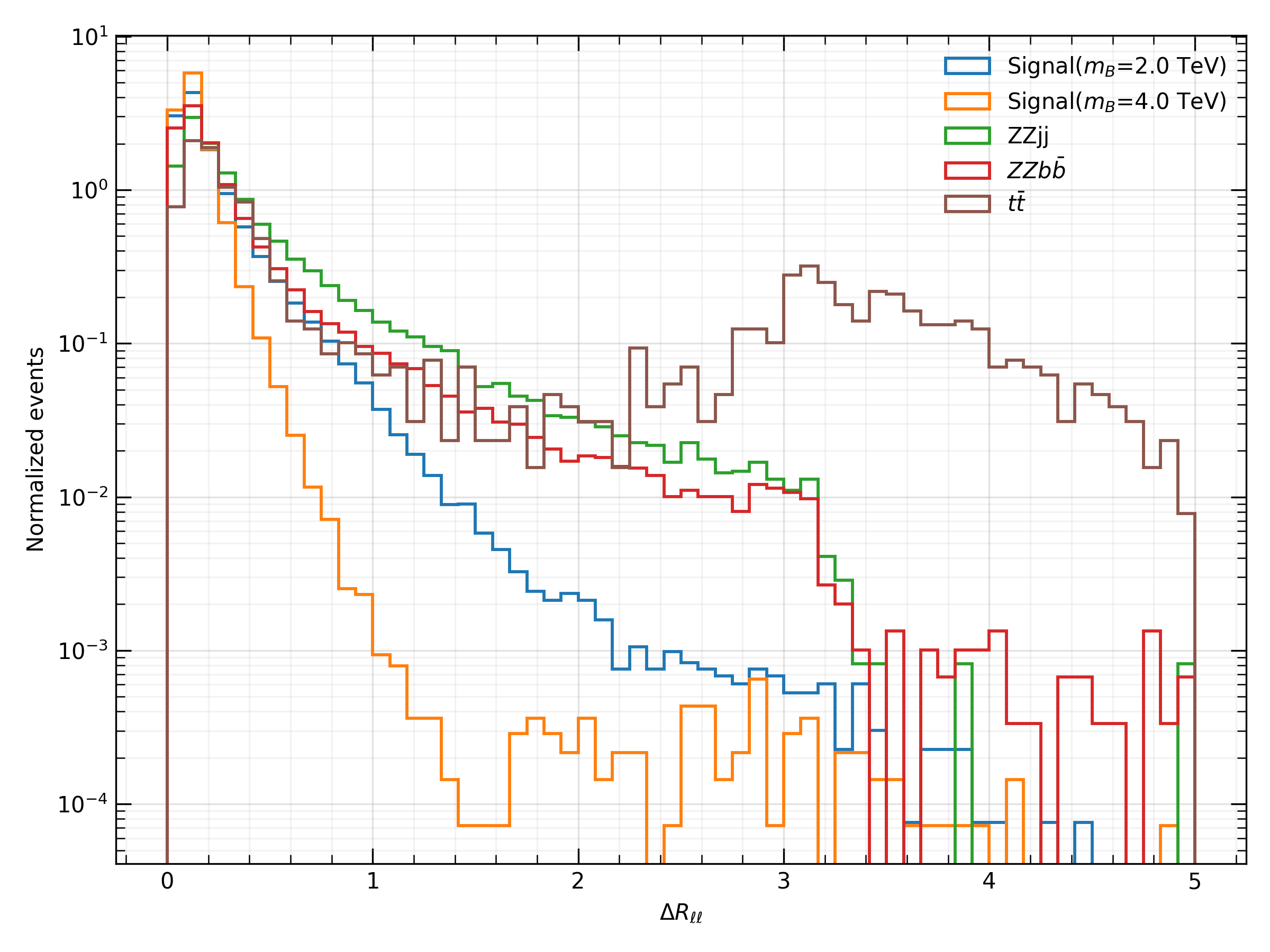}
    \par\small (c) Angular separation distribution between the dilepton system, $\Delta R_{\ell\ell}$
\end{minipage}
\hfill
\begin{minipage}{0.48\textwidth}
    \centering
    \includegraphics[width=\textwidth]{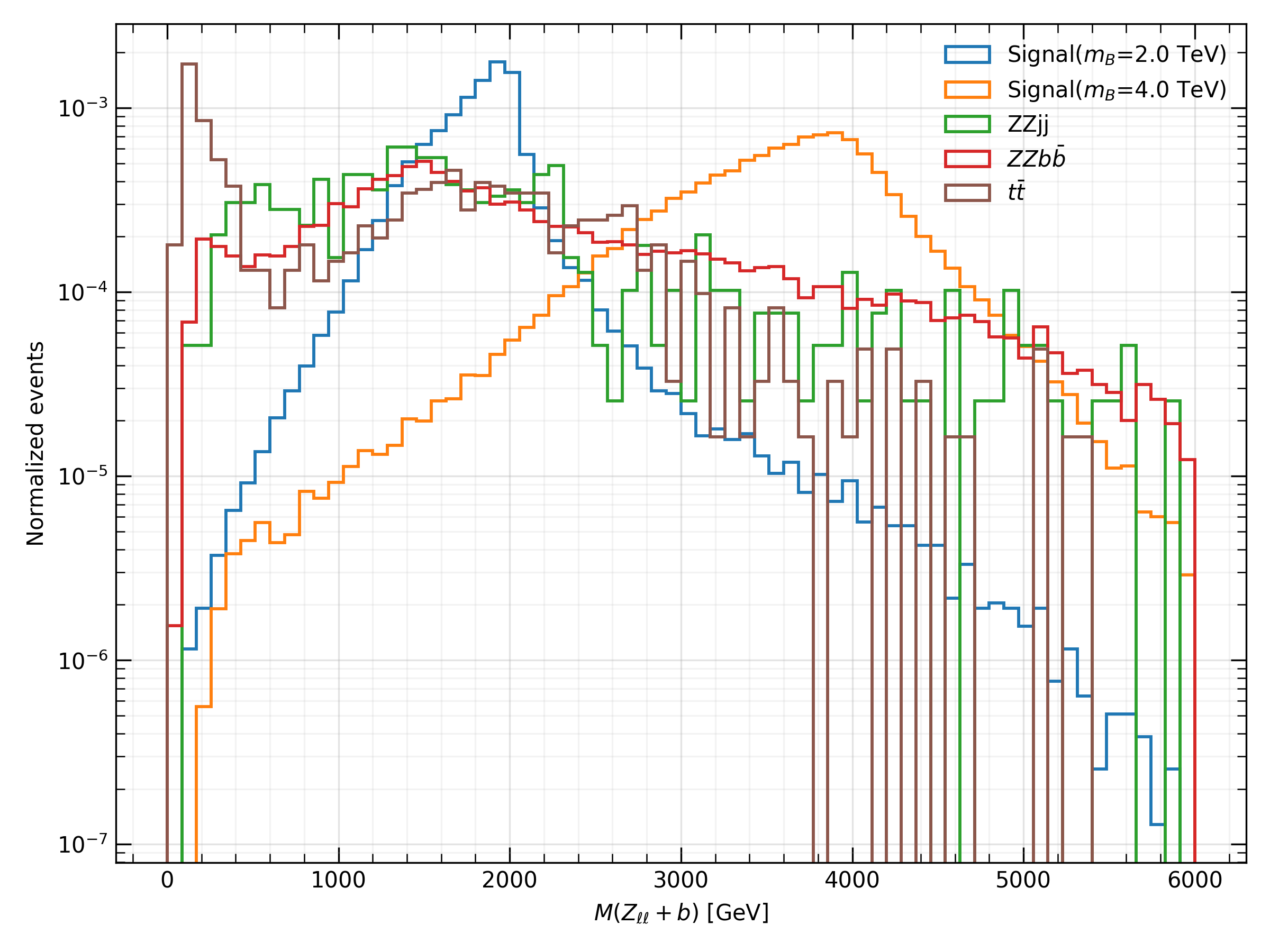}
    \par\small (d) Distribution of the reconstructed leptonic VLQ-$B$ candidate mass,
$M(Z_{\ell\ell}+b)$, for the signal and background processes.
\end{minipage}
\caption{Kinematic distributions of the main observables used in the analysis for the signal and background processes after the basic event selection.}
\label{fig:kinematic_overview}
\end{figure*}

The kinematic distributions used to motivate the event selection are shown in Fig.~\ref{fig:kinematic_overview}. Figure~\ref{fig:kinematic_overview}(a) presents the transverse momentum of the leading $b$-tagged jet, $p_T(b_1)$. As expected from the decay of a heavy VLQ-$B$ resonance, the signal samples lead to substantially harder $b$-jet spectra than the SM backgrounds. The $ZZjj$ and $ZZb\bar{b}$ backgrounds are mostly confined to the low-$p_T$ region, whereas the $t\bar{t}$ contribution develops a moderate tail. This feature motivates the requirement on $p_T(b_1)$ used in the event selection.

Figure~\ref{fig:kinematic_overview}(b) shows the scalar sum of jet transverse momenta, $H_T$. The signal distributions extend to large $H_T$ values due to the energetic hadronic activity produced in the decay chain of the pair-produced VLQ-$B$ quarks. By contrast, the background distributions decrease rapidly in the high-$H_T$ region, making this observable particularly effective for suppressing SM contributions.

The angular separation between the two selected leptons, $\Delta R_{\ell\ell}$, is displayed in Fig.~\ref{fig:kinematic_overview}(c). Signal events tend to populate small $\Delta R_{\ell\ell}$ values, reflecting the boosted leptonic $Z$ boson produced in the heavy-quark decay. The backgrounds are more broadly distributed and are shifted toward larger angular separations, which supports the use of a boosted dilepton selection.

Finally, Fig.~\ref{fig:kinematic_overview}(d) presents the reconstructed invariant mass of the leptonic VLQ-$B$ candidate, $M(Z_{\ell\ell}+b)$. The signal distributions for the two benchmark points, $M_B=2~\mathrm{TeV}$ and $M_B=4~\mathrm{TeV}$, peak in the high-mass region and remain well separated from the SM backgrounds, which are concentrated at lower invariant masses. This observable therefore provides a direct handle on the heavy-quark mass scale and serves as an important consistency check of the signal topology.

Based on these features, we define the event selection as follows, aiming to retain the signal efficiency while reducing the dominant SM backgrounds.
\begin{itemize}
\item \textbf{Basic cuts:} Leptons and jets are required to satisfy $p^\ell_T > 30~\mathrm{GeV}$, $|\eta^\ell| < 2.5$ and $p^j_T > 30~\mathrm{GeV}$, $|\eta^j| < 2.5$.
\item \textbf{Cut~1:} Events are required to contain at least one opposite-sign same-flavor dilepton pair together with at least four jets and at least two $b$-tagged jets, $N(\ell^+\ell^-)_\mathrm{OS-SF} \geq 1$, $N(j) \geq 4$, $N(b) \geq 2$.
\item \textbf{Cut~2:} The transverse momentum of the leading $b$-tagged jet is required to satisfy $p_T(b_1) > 200~\mathrm{GeV}$.
\item \textbf{Cut~3:} To exploit the boosted dilepton topology of the signal events, the angular separation between the leptons is required to satisfy $\Delta R_{\ell\ell} < 1.0$.
\item \textbf{Cut~4:} Large hadronic activity is imposed through the scalar sum of the jet transverse momenta, $H_T > 2000~\mathrm{GeV}$.
\end{itemize}
The impact of the successive selection criteria on the signal and background cross sections is summarised in Table~\ref{tab:cutflow}.

Among the applied selection criteria, Cut~1 provides the most significant background rejection, reducing the $ZZjj$ background by nearly two orders of magnitude while retaining a large fraction of the signal events. The subsequent kinematic cuts (Cut~2, Cut~3, and Cut~4) have a negligible effect on the signal, indicating that the signal events naturally satisfy these requirements due to the large mass scale of the VLQ-$B$ quark. In contrast, these cuts further suppress the remaining background contributions, particularly the $t\bar{t}$ process.

The $ZZb\bar{b}$ background exhibits the highest final efficiency among the SM processes, as its final-state topology most closely resembles that of the signal. The $t\bar{t}$ background, despite its large initial cross section, is strongly suppressed by Cut~1 because the typical $t\bar{t}\to W^+bW^-\bar b$ topology does not efficiently satisfy the simultaneous requirements of an OS--SF dilepton pair, at least four jets, and two $b$-tagged jets. The residual contribution is then further reduced by the large $H_T$ threshold in Cut~4. The final signal selection efficiencies are approximately 46\% and 52\% for $M_B = 2~\mathrm{TeV}$ and $M_B = 4~\mathrm{TeV}$, respectively, with the higher-mass benchmark yielding a larger efficiency owing to the more boosted nature of its decay products.

\section{Statistical Analysis and Sensitivity Results}
\label{sec:analysis}
As it is well known, statistical significance calculations are required to quantitatively evaluate the effect of the distinction between the signal and SM backgrounds on the discovery potentials in terms of experimental sensitivity. Accordingly, the expression based on the Asimov statistic given by Cowan et al.\ is adopted to quantify the discovery sensitivities~\cite{Cowan2011Asymptotic}.

\begin{figure*}[t]
\centering
\begin{subfigure}{0.49\textwidth}
    \centering
    \includegraphics[width=\linewidth]{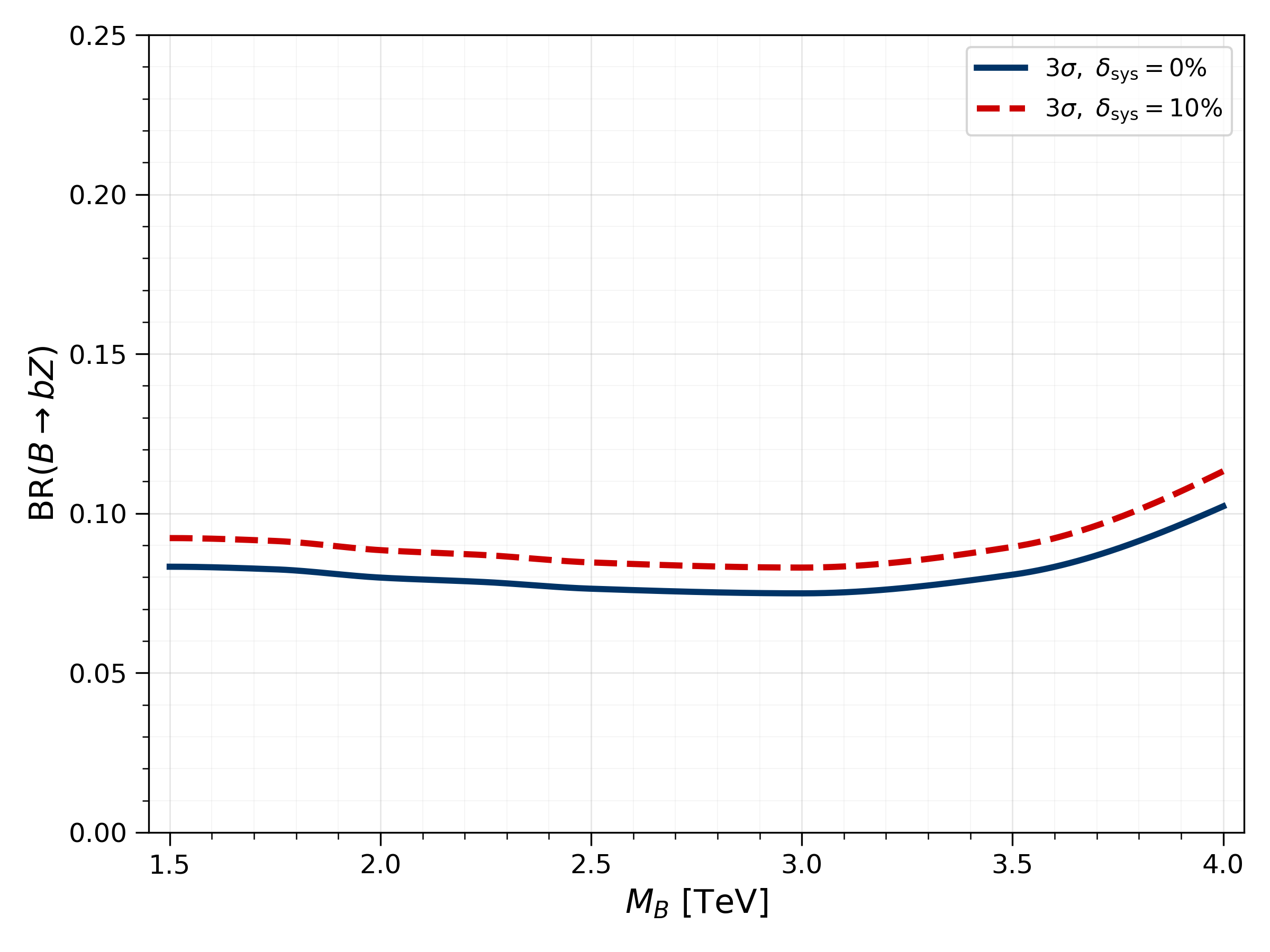}
    \caption{$3\sigma$ evidence sensitivity.}
    \label{fig:br_mass_3sigma}
\end{subfigure}
\hfill
\begin{subfigure}{0.49\textwidth}
    \centering
    \includegraphics[width=\linewidth]{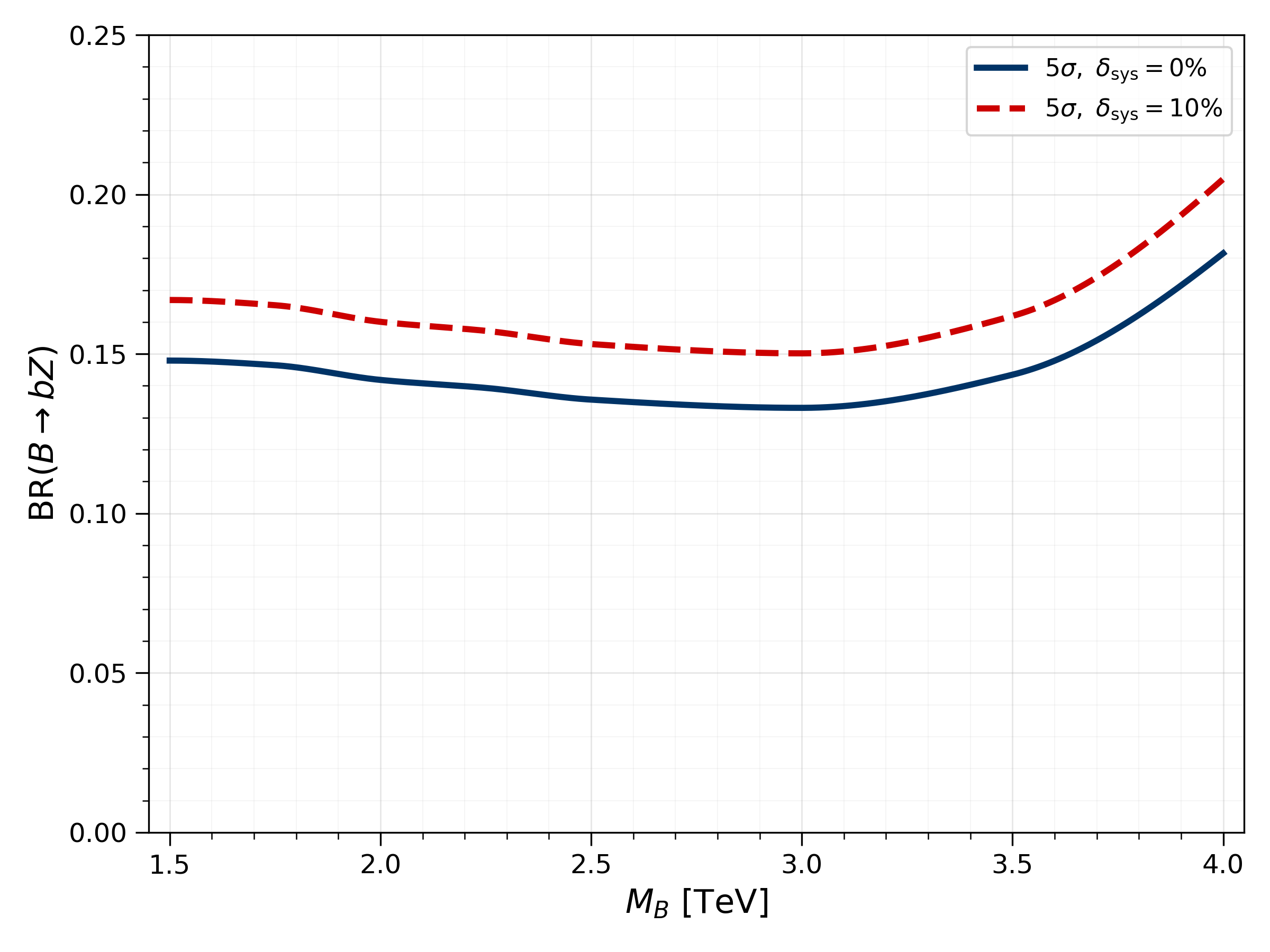}
    \caption{$5\sigma$ discovery sensitivity.}
    \label{fig:br_mass_5sigma}
\end{subfigure}
\caption{
Required branching ratio $\mathrm{BR}(B\to bZ)$ as a function of the VLQ-$B$ mass
for an integrated luminosity of $\mathcal{L}=10~\mathrm{ab}^{-1}$.
Solid (dashed) curves correspond to
$\delta_{\rm sys}=0\%$ ($10\%$) background systematic uncertainty.
}
\label{fig:br_mass}
\end{figure*}

\begin{figure*}[t]
\centering
\begin{subfigure}{0.49\textwidth}
    \centering
    \includegraphics[width=\linewidth]{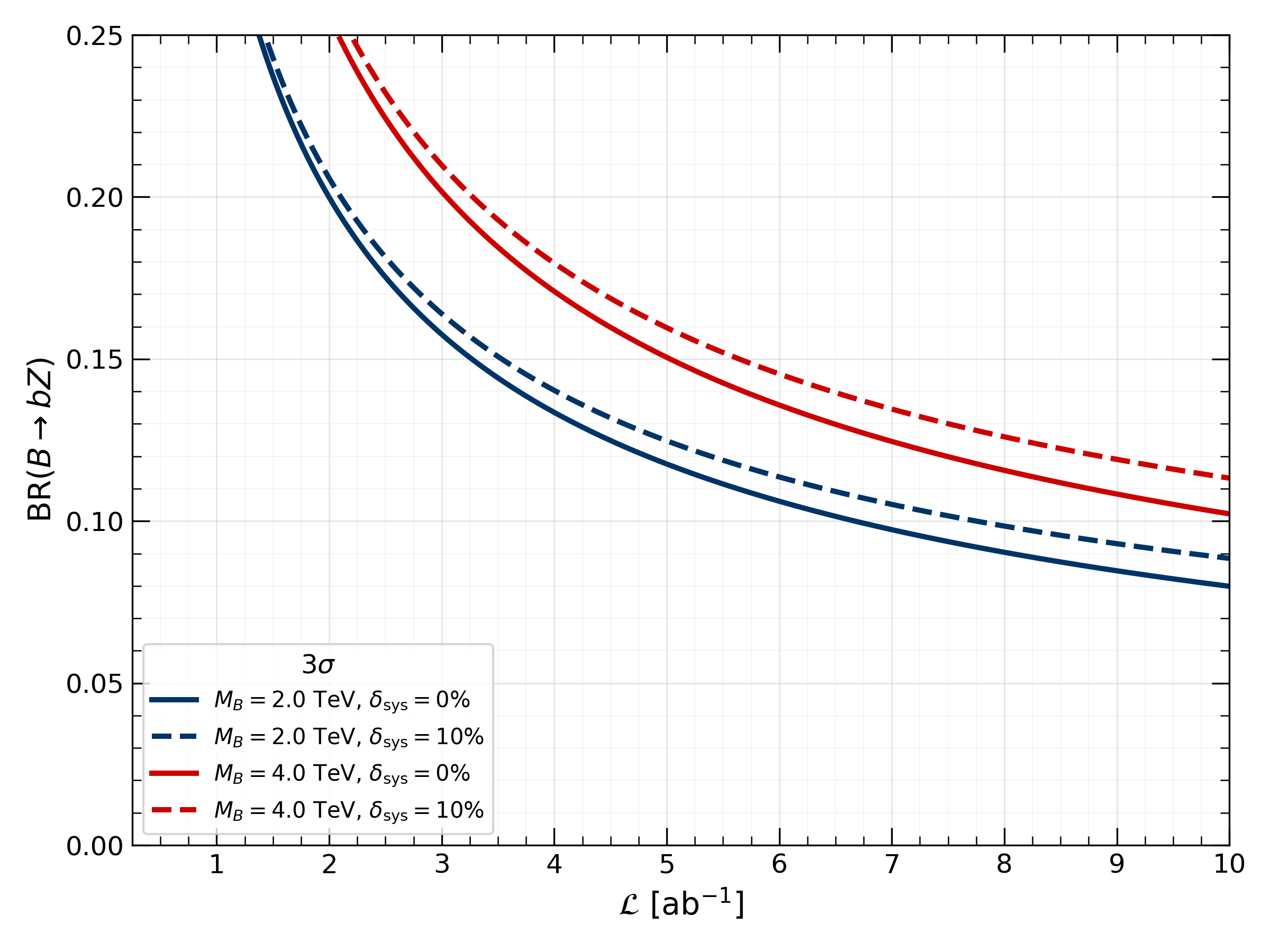}
    \caption{$3\sigma$ evidence sensitivity as a function of the integrated luminosity.}
    \label{fig:br_lumi_3sigma}
\end{subfigure}
\hfill
\begin{subfigure}{0.49\textwidth}
    \centering
    \includegraphics[width=\linewidth]{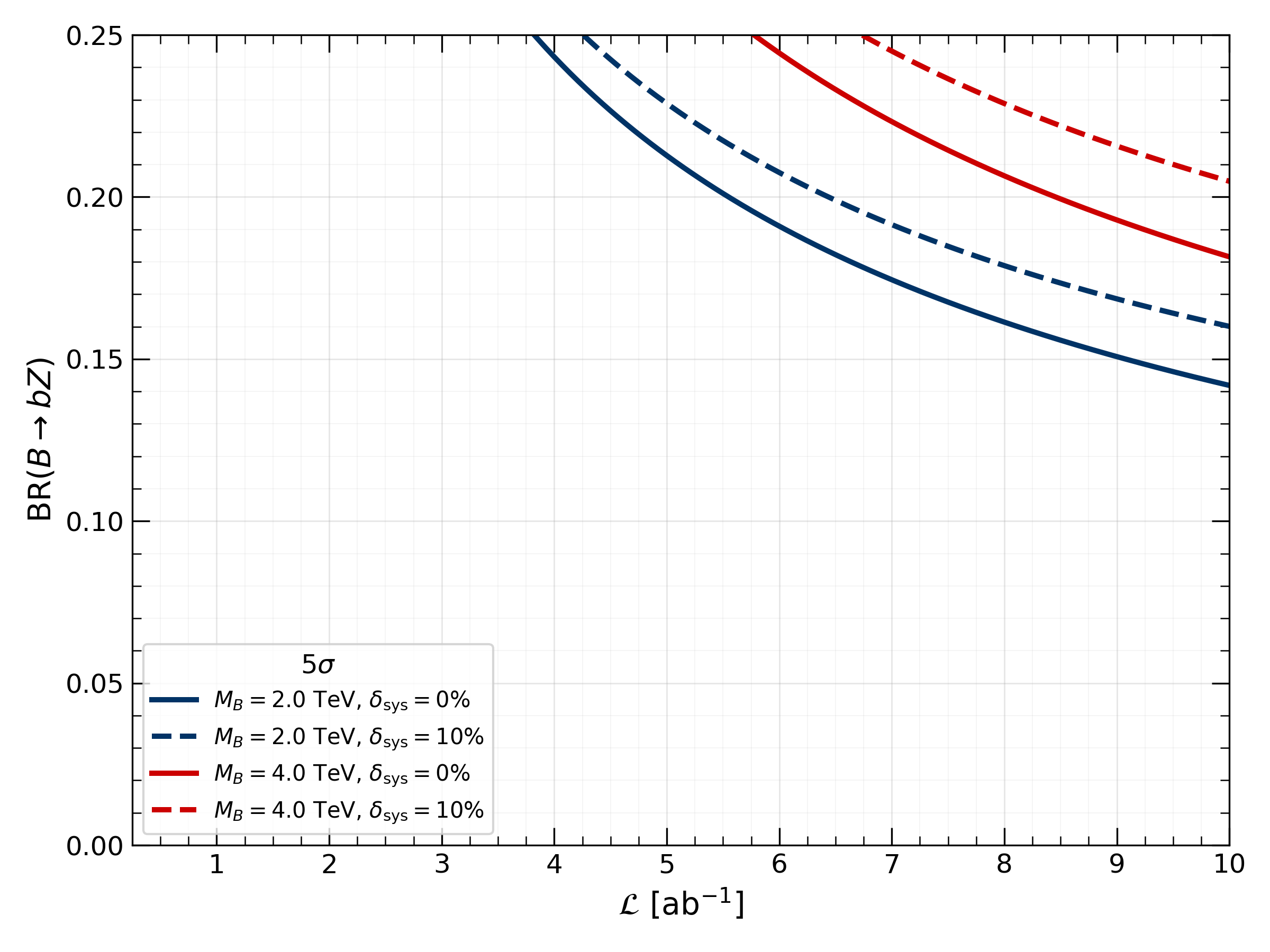}
    \caption{$5\sigma$ discovery sensitivity as a function of the integrated luminosity.}
    \label{fig:br_lumi_5sigma}
\end{subfigure}
\caption{
Required branching ratio $\mathrm{BR}(B\to bZ)$ as a function of the integrated luminosity for
$M_B=2$ and $4~\mathrm{TeV}$ benchmark scenarios.
Solid (dashed) curves correspond to
$\delta_{\rm sys}=0\%$ ($10\%$) background systematic uncertainty.
}
\label{fig:br_lumi}
\end{figure*}

The signal and background event numbers are defined as $N_{S,B}=\sigma_{S,B}\times\mathcal{L}$, where $\sigma_S$ and $\sigma_B$ correspond to the cross sections after the applied cuts for the signal and the total background, respectively, and $\mathcal{L}$ represents the integrated luminosity. When the background systematic uncertainty is denoted by $\delta$, the discovery significance is defined as
\begin{multline}
Z_{\rm disc}
=
\Bigg\{
2 \Bigg[
(s+b)\,
\ln\!\left(
\frac{(s+b)\,(1+\delta^2 b)}
{b + \delta^2 b\,(s+b)}
\right)
\\
-\frac{1}{\delta^2}\,
\ln\!\left(
1 + \frac{\delta^2 s}{1+\delta^2 b}
\right)
\Bigg]
\Bigg\}^{1/2}
\label{eq:Zdisc_delta}
\end{multline}
which reduces to the following expression in the limit $\delta\to0$:
\begin{equation}
Z_{\rm disc}=
\sqrt{2\left[(s+b)\ln\left(1+\frac{s}{b}\right)-s\right]}.
\label{eq:Zdisc_delta0}
\end{equation}
The sensitivity results, expressed in terms of the minimum required branching ratio $\mathrm{BR}(B\to bZ)$, are presented in Fig.~\ref{fig:br_mass} as a function of the VLQ-$B$ mass and in Fig.~\ref{fig:br_lumi} as a function of the integrated luminosity. In both figures, solid curves correspond to the case of no background systematic uncertainty ($\delta_\mathrm{sys}=0\%$), while dashed curves represent the results obtained with a 10\% systematic uncertainty ($\delta_\mathrm{sys}=10\%$). Both the $5\sigma$ discovery and $3\sigma$ evidence thresholds are presented to provide a complete picture of the achievable sensitivity.

As mentioned in the Introduction, the Goldstone boson equivalence theorem predicts the asymptotic branching pattern $\mathrm{BR}(B\to Wt):\mathrm{BR}(B\to bZ):\mathrm{BR}(B\to Hb)\simeq2:1:1$ for a singlet VLQ-$B$ in the large-mass limit. Accordingly, we adopt $\mathrm{BR}(B\to bZ)=25\%$ as the benchmark value throughout this analysis. Nevertheless, to express the sensitivity in a more model-independent form, the branching ratio can also be treated as a free effective parameter. This approach is particularly relevant in scenarios where additional non-standard decay modes, such as $B\to bS$ with a new scalar or pseudoscalar state $S$~\cite{Banerjee2024TripletVLQ}, are kinematically allowed. In such cases, the total branching ratio into the standard channels $B\to Wt$, $B\to Zb$, and $B\to Hb$ is reduced, leading to a smaller effective value of $\mathrm{BR}(B\to bZ)$. The sensitivity curves shown in Figs.~\ref{fig:br_mass} and~\ref{fig:br_lumi} can therefore also be interpreted as a probe of scenarios with diluted standard VLQ decay modes.

With this interpretation in mind, Fig.~\ref{fig:br_mass} shows that, for a fixed integrated luminosity of $\mathcal{L}=10~\mathrm{ab}^{-1}$, the minimum required branching ratio remains relatively stable over the lower part of the mass range and increases toward higher VLQ-$B$ masses. This behaviour reflects the interplay between the decreasing pair-production cross section and the changing selection efficiency across the scanned mass range. For the $5\sigma$ discovery threshold, the minimum required branching ratios at the two benchmark mass points are $\mathrm{BR}(B\to bZ)\approx14.2\%$ and $\approx16.3\%$ for $M_B=2~\mathrm{TeV}$ with $\delta_\mathrm{sys}=0\%$ and $10\%$, respectively, and $\mathrm{BR}(B\to bZ)\approx18.2\%$ and $\approx20.5\%$ for $M_B=4~\mathrm{TeV}$ under the same systematic-uncertainty assumptions. For the $3\sigma$ evidence threshold, the corresponding values are $\approx7.8\%$ and $\approx9.0\%$ at $M_B=2~\mathrm{TeV}$, and $\approx10.1\%$ and $\approx11.6\%$ at $M_B=4~\mathrm{TeV}$. These values remain below the singlet benchmark expectation $\mathrm{BR}(B\to bZ)=25\%$, indicating that the benchmark scenario can be tested over the considered mass range at both the evidence and discovery levels. Including a 10\% background systematic uncertainty increases the required branching ratio by about two percentage points, corresponding to a reduction in the discovery reach of roughly $100~\mathrm{GeV}$.

Assuming the singlet benchmark value $\mathrm{BR}(B\to bZ)=0.25$, a $5\sigma$ discovery can be achieved up to approximately $M_B\simeq4.3~\mathrm{TeV}$ for $\mathcal{L}=10~\mathrm{ab}^{-1}$ in the absence of systematic uncertainties. The reach is reduced slightly to $M_B\simeq4.2~\mathrm{TeV}$ when a 10\% background systematic uncertainty is included.

Similarly, Fig.~\ref{fig:br_lumi} illustrates the luminosity dependence of the sensitivity for the two benchmark mass points. The minimum required branching ratio decreases as the integrated luminosity increases, with the most pronounced change occurring in the lower-luminosity region and a more gradual decrease at higher luminosities. As expected, the $M_B=4~\mathrm{TeV}$ benchmark requires a larger branching ratio than the $M_B=2~\mathrm{TeV}$ case, reflecting the smaller pair-production cross section at higher masses. The inclusion of a 10\% background systematic uncertainty shifts the curves upward, indicating a weaker sensitivity compared with the purely statistical case. Both the $5\sigma$ discovery and $3\sigma$ evidence thresholds are shown, with the latter corresponding to a statistically significant indication of a possible signal rather than a formal discovery claim.

For the $5\sigma$ discovery of a VLQ-$B$ with $M_B=2~\mathrm{TeV}$, the minimum required integrated luminosities are $3.82~\mathrm{ab}^{-1}$ and $4.26~\mathrm{ab}^{-1}$ for $\delta_\mathrm{sys}=0\%$ and $10\%$, respectively. The corresponding $3\sigma$ evidence thresholds are reached at $1.38~\mathrm{ab}^{-1}$ and $1.43~\mathrm{ab}^{-1}$. For the heavier benchmark point, $M_B=4~\mathrm{TeV}$, the $5\sigma$ luminosity requirements increase to $5.77~\mathrm{ab}^{-1}$ and $6.73~\mathrm{ab}^{-1}$, while the $3\sigma$ evidence level is reached at $2.08~\mathrm{ab}^{-1}$ and $2.19~\mathrm{ab}^{-1}$ under the same systematic-uncertainty assumptions. These values remain below the target integrated luminosity of $10~\mathrm{ab}^{-1}$ for a $\sqrt{s}=10~\mathrm{TeV}$ muon collider. This suggests that both benchmark masses remain within the projected $5\sigma$ discovery reach, even after including a 10\% background systematic uncertainty.

For instance, a 100~TeV FCC-hh study of single VLQ-$B$ production in the $B\to tW$ channel reported an exclusion reach beyond $M_B=5~\mathrm{TeV}$ for the benchmark coupling $g^\ast=0.2$. In the present work, a $5\sigma$ discovery reach of $M_B\simeq4.2$--$4.3~\mathrm{TeV}$ is obtained for VLQ-$B$ pair production in the $bZ$ channel at a $10~\mathrm{TeV}$ muon collider. Although the results of these two studies correspond to different production mechanisms and final states, they indicate that the muon-collider search considered here probes a mass scale close to that reached in the FCC-hh benchmark, while offering a complementary lepton-collider environment. Furthermore, compared with 3~TeV CLIC pair-production studies, the higher collision energy considered in this work extends the accessible VLQ-$B$ mass range well beyond the kinematic limit of approximately $M_B\lesssim1.5~\mathrm{TeV}$.

Consequently, the $10~\mathrm{TeV}$ muon collider considered in this work extends the accessible VLQ-$B$ mass range well beyond the reach of 3~TeV CLIC pair-production studies and provides a complementary probe to future hadron-collider searches. These findings indicate that a high-energy muon collider can provide enhanced sensitivity to vector-like $B$ quarks in the multi-TeV mass range.

\section{Conclusion}
\label{sec:conclusion}

We have investigated the pair production of singlet vector-like $B$ quarks at a $\sqrt{s}=10~\mathrm{TeV}$ muon collider via
$\mu^+\mu^- \to B\bar{B} \to (Zb)(Z\bar{b}) \to \ell^+\ell^- b\, jj\bar{b}$,
exploiting the boosted $Zb$ final-state topology. A cut-based analysis targeting hard leading-$b$-jet transverse momentum, small dilepton angular separation, and large hadronic activity achieves signal efficiencies of $46$--$52\%$ while strongly suppressing the dominant $ZZjj$, $ZZb\bar{b}$, and $t\bar{t}$ backgrounds.

For the singlet benchmark $\mathrm{BR}(B\to bZ)=0.25$ and
$\mathcal{L}=10~\mathrm{ab}^{-1}$, a $5\sigma$ discovery reach of
$M_B\simeq 4.3~\mathrm{TeV}$ is obtained, reduced to
$M_B\simeq 4.2~\mathrm{TeV}$ when a $10\%$ background systematic uncertainty is included. This sensitivity significantly exceeds the current LHC exclusion limits and the projected reach of lower-energy lepton colliders. The $5\sigma$ discovery luminosity requirements are
$3.82~\mathrm{ab}^{-1}$ and $5.77~\mathrm{ab}^{-1}$ for
$M_B=2$ and $4~\mathrm{TeV}$, respectively, both well within the design luminosity of a $10~\mathrm{TeV}$ muon collider.

These results show that a high-energy muon collider provides a clean and sensitive environment for probing vector-like $B$ quarks in the multi-TeV regime, establishing the boosted $bZ$ channel as a promising discovery mode for future energy-frontier searches.

\section*{Acknowledgements}
This work was supported by the Scientific Research Projects Coordination Unit of
Zonguldak Bülent Ecevit University under Project No.~2025-22794455-04.

\section*{Data Availability Statement}
No new data were generated in this study.

\section*{Code Availability Statement}
No new code was generated in this study.

\begin{table*}[t]
\centering
\caption{Cut-flow table for the signal and background processes after successive selection criteria.
The entries correspond to the cross sections (in fb) after each selection step.}
\label{tab:cutflow}
\begin{tabular}{lcccccc}
\hline\hline
Cuts &
Signal ($M_B=2.0$ TeV) &
Signal ($M_B=4.0$ TeV) &
ZZjj &
ZZbb &
$t\bar t$
\\
\hline
Initial
& $9.215\times10^{-3}$
& $6.286\times10^{-3}$
& $7.540\times10^{-3}$
& $4.160\times10^{-3}$
& $4.693\times10^{-1}$
\\
Basic cuts
& $8.688\times10^{-3}$
& $6.080\times10^{-3}$
& $3.254\times10^{-3}$
& $2.001\times10^{-3}$
& $3.651\times10^{-1}$
\\
Cut 1
& $4.274\times10^{-3}$
& $3.289\times10^{-3}$
& $3.974\times10^{-5}$
& $6.568\times10^{-4}$
& $1.646\times10^{-3}$
\\
Cut 2
& $4.265\times10^{-3}$
& $3.289\times10^{-3}$
& $3.885\times10^{-5}$
& $6.245\times10^{-4}$
& $1.635\times10^{-3}$
\\
Cut 3
& $4.207\times10^{-3}$
& $3.286\times10^{-3}$
& $3.381\times10^{-5}$
& $5.818\times10^{-4}$
& $1.064\times10^{-3}$
\\
Cut 4
& $4.205\times10^{-3}$
& $3.286\times10^{-3}$
& $3.285\times10^{-5}$
& $5.532\times10^{-4}$
& $1.018\times10^{-3}$
\\
Efficiency (\%)
& 45.63
& 52.27
& 0.44
& 13.30
& 0.22
\\
\hline\hline
\end{tabular}
\end{table*}

\FloatBarrier
\bibliographystyle{unsrt}
\bibliography{references}
\end{document}